\documentstyle[11pt,newpasp,twoside,epsfig]{article}
\reversemarginpar

\def\edcomment#1{\iffalse\marginpar{\raggedright\sl#1\/}\else\relax\fi}

\def\eg{{e.g., }}
\def\ie{{i.e., }}
\def\etal{{et al., }}
\def\et{{et al. }}
\def\etc{{etc.}}

\def\'{^{\prime}}

\def\hmpc{{\, {\rm h}^{-1}~\rm Mpc}}

\def\mpc{{\rm~Mpc}}

\def\spose#1{\hbox to 0pt{#1\hss}}
\def\lta{\mathrel{\spose{\lower 3pt\hbox{$\mathchar"218$}}
     \raise 2.0pt\hbox{$\mathchar"13C$}}}
\def\gta{\mathrel{\spose{\lower 3pt\hbox{$\mathchar"218$}}
     \raise 2.0pt\hbox{$\mathchar"13E$}}}
\def\ge{\mathrel{\spose{\lower 3pt\hbox{$-$}}
     \raise 2.0pt\hbox{$\mathchar"13E$}}}
\def\le{\mathrel{\spose{\lower 3pt\hbox{$-$}}
     \raise 2.0pt\hbox{$\mathchar"13C$}}}

\begin{document}

\title{CMB Analysis of Boomerang \& Maxima \\
\& the Cosmic Parameters $\{\Omega_{tot},\Omega_b{\rm
h}^2,\Omega_{cdm}{\rm h}^2, \Omega_\Lambda,n_s \}$}

\newcommand{\cita}{Canadian Institute for Theoretical Astrophysics,
  University of Toronto, Canada}
\newcommand{\cfpa}{Center for Particle
  Astrophysics, University of California, Berkeley, CA, USA}
\newcommand{\caltech}{California Institute of  Technology, Pasadena, CA, USA}
\newcommand{\spacescience}{Space Sciences Laboratory, University of
  California, Berkeley, CA, USA}
\newcommand{\ucsb}{Department of Physics, University of California, Santa
  Barbara, CA, USA}
\newcommand{\ENEA}{ENEA Centro Ricerche di Frascati,
  Via E. Fermi 45, 00044 Frascati, Italy}
\newcommand{\ucbphysics}{Department of Physics, University of California,
  Berkeley CA, USA}
\newcommand{\ucbastro}{Department of Astronomy, University of California,
  Berkeley CA, USA}
\newcommand{\lasapienza}{Dipartimento di Fisica, Universita' La
  Sapienza, Roma, Italy}

\author{J. Richard~Bond$^{1}$, P.A.R.~Ade$^{2}$, A.~Balbi$^{3,4}$, J.J~Bock$^{5}$,
J.~Borrill$^{6}$, A.~Boscaleri$^{7}$, K.~Coble$^{8}$,
B.P.~Crill$^{9}$, P.~de~Bernardis$^{10}$, P.~Farese$^{8}$,
P.G.~Ferreira$^{11}$, K.~Ganga$^{9,12}$, M.~Giacometti$^{10}$,
S.~Hanany$^{13}$, E.~Hivon$^{9}$, V.V.~Hristov$^{9}$,
A.~Iacoangeli$^{10}$, A.H.~Jaffe$^{3}$, A.E.~Lange$^{9}$,
A.T.~Lee$^{3}$, L.~Martinis$^{14}$, S.~Masi$^{10}$,
P.D.~Mauskopf$^{15}$, A.~Melchiorri$^{10}$, T.~Montroy$^{8}$,
C.B.~Netterfield$^{16}$, S.~Oh$^{3}$, E.~Pascale$^{7}$,
F.~Piacentini$^{10}$, D.~Pogosyan$^{1}$, S.~Prunet$^{1}$,
B.~Rabii$^{9}$, S.~Rao$^{17}$, P.L.~Richards$^{3}$,
G.~Romeo$^{17}$, J.E.~Ruhl$^{8}$, F.~Scaramuzzi$^{14}$,
D.~Sforna$^{10}$, K.~Sigurdson$^{1,9}$, G.F.~Smoot$^{3}$,
R.~Stompor$^{3}$, C.D.~Winant$^{3}$, J.H.P.~Wu$^{3}$} \affil{1.
\cita 2. Queen Mary and Westfield College, London, UK 3. \cfpa 4.
Dipartimento di Fisica, Universit\`a Tor Vergata, Roma, Italy
 5. Jet Propulsion Laboratory, Pasadena, CA, USA, 6. National
Energy Research Scientific Computing Center, LBNL,
  Berkeley, CA, USA, 7. IROE-CNR, Firenze, Italy,
8. \ucsb 9. \caltech 10. \lasapienza 11. Astrophysics, University
of Oxford, NAPL, Keble Road, OX2 6HT, UK 12. Physique
Corpusculaire et Cosmologie, Coll\`ege de France, 11 Place
Marcelin Berthelot, 75231 Paris Cedex 05, France 13. School of
Physics and Astronomy, University of Minnesota/Twin Cities,
Minneapolis, MN, USA 14. \ENEA 15. University of Wales, Cardiff,
UK, CF24 3YB 16. Departments of Physics and Astronomy, University
of Toronto, Canada 17. Istituto Nazionale di Geofisica, Roma,
Italy\\
{\tt \small CITA-2000-65, in Proc. IAU Symposium 201 (PASP), eds.
A. Lasenby, A. Wilkinson\normalsize}
}

\begin{abstract}
We show how estimates of parameters characterizing inflation-based
theories of  structure formation localized over the past year when
large scale structure (LSS) information from galaxy and cluster
surveys was combined with the rapidly developing cosmic microwave
background (CMB) data,  especially from  the recent Boomerang and
Maxima balloon experiments. All current CMB data plus a relatively
weak  prior probability on the Hubble constant, age and LSS points
to little mean curvature ($\Omega_{tot} = 1.08\pm 0.06$) and
nearly scale invariant initial fluctuations ($n_s =1.03 \pm
0.08$), both predictions of (non-baroque) inflation theory. We
emphasize the role that degeneracy among parameters in the $L_{pk}
= 212\pm 7$  position of the (first acoustic) peak plays in
defining the $\Omega_{tot}$ range upon marginalization over other
variables. Though the CDM density is in the expected range
($\Omega_{cdm}{\rm h}^2=0.17 \pm 0.02$), the baryon density $
\Omega_b {\rm h}^2 = 0.030\pm 0.005$ is somewhat above the
independent $0.019\pm 0.002$ nucleosynthesis estimates. CMB+LSS
gives independent evidence for dark energy ($\Omega_\Lambda = 0.66
\pm 0.06$) at the same level as from supernova (SN1) observations,
with a phenomenological quintessence equation of state limited by
SN1+CMB+LSS to $w_Q < -0.7$ cf. the $w_Q$=$-1$ cosmological
constant case.
\end{abstract}

\section{CMB Analysis of Primary Anisotropies}

\noindent {\bf Experiments and Bandpowers:} Anisotropies at the
$30 \mu K$ level at low multipoles  revealed by COBE in 1992 were
augmented at higher $\ell$ in some 19 other experiments, some with
a comparable number of resolution elements to the 600 or so for
COBE, most with many fewer. A list of these experiments to April
1999 with associated bandpowers is given in Bond, Jaffe and Knox
(2000 [BJK00]). The anisotropy picture dramatically improved this
past year, as results were announced first in summer 99 from the
ground-based TOCO experiment in Chile (Miller \et 2000), then in
November 99 from Boomerang-NA, the North American test flight
(Mauskopf et 1999). These two additions improved peak localization
and gave evidence for $\Omega_{tot}\sim 1$. Then in April 2000,
results  from the first CMB long duration balloon (LDB) flight,
were announced (de Bernardis \et 2000), followed in May 2000 by
results from the night flight of Maxima (Hanany \et 2000).
Boomerang's best resolution was $10^\prime$, about 40 times better
than that of COBE, with tens of thousands of resolution elements.
Maxima had a similar resolution but covered an order of magnitude
less sky. Fig.~1 shows the 150A GHz Boomerang-LDB map and the
Wiener-filtered Maxima-1, to scale. The de Bernardis \et (2000)
maps at 90 and 220 GHz show the same spatial features as this  150
GHz one, with the overall intensities falling precisely on the CMB
blackbody curve.  The Toco, Boomerang and Maxima experiments are
described elsewhere in these proceedings. They were designed to
reveal the {\it primary} anisotropies of the CMB, those which can
be calculated using linear perturbation theory. Fig.~1 shows the
temperature power spectra for Boomerang,  Maxima and prior-CMB
data (Boomerang-NA+TOCO+April 99) are in good agreement. Sketching
the impact of these new results on cosmic parameter estimation
(Lange \et 2000 [Let00], Jaffe \et 2000 [Jet00]) is the goal of
this paper. Space constraints preclude adequate referencing here,
but these are given in the Boomerang (Let00) and Maxima+Boomerang
(Jet00) parameter estimation papers (see also Bond 1996, [B96],
for other references).

\begin{figure}
\vspace{-15pt}
\centerline{\hspace{30pt}\epsfig{file=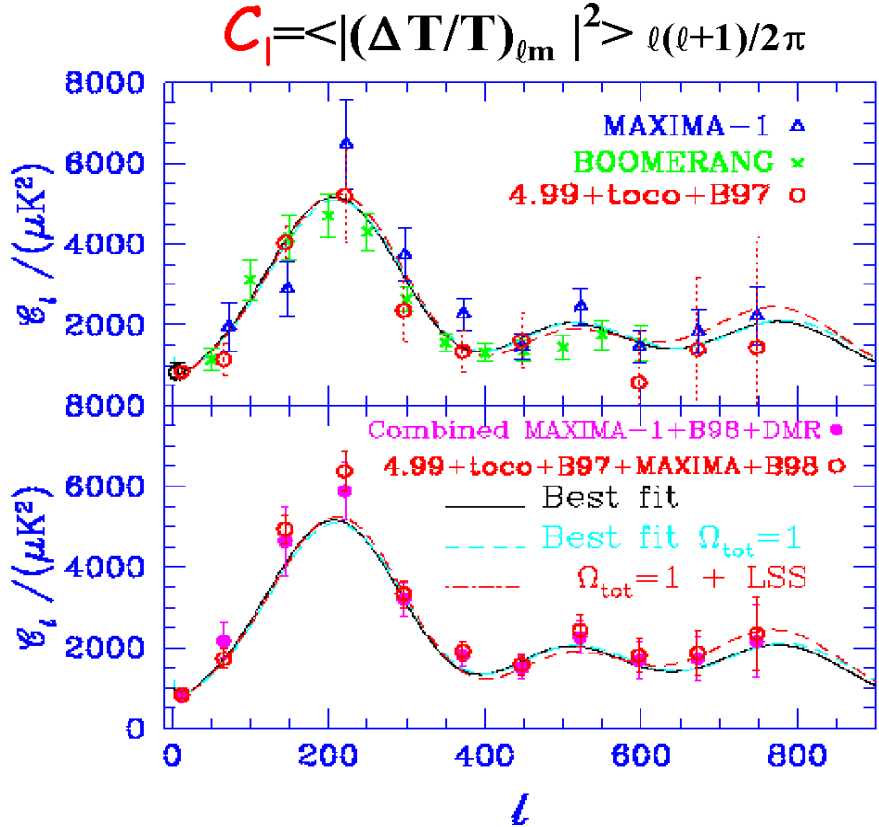,height=4.5in}}
\vspace{-7pt} \centerline{\epsfig{file=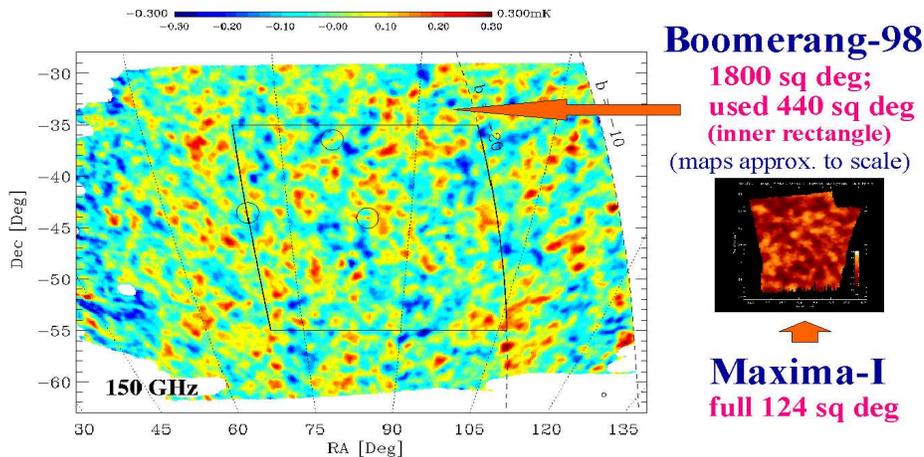,height=3.3in}}
\vspace{-12pt} \caption{\small The top figure shows ${\cal
C}_\ell$ grouped in bandpowers for Boomerang-LDB (crosses),
Maxima-I (triangles) and prior-CMB experiments
(TOCO+Boomerang-NA+"April 99", squares). The lower panel contrasts
the optimally-combined power spectra for Boomerang+Maxima+DMR
(squares) with that for Boomerang+Maxima+prior-CMB (circles),
showing the prior experiments do not move ${\cal C}_\ell$ very
much. Best-fit models for arbitrary $\Omega_{tot}$ and for
$\Omega_{tot}$=1 are shown in both panels. The Boomerang 150A GHz
map (\ie for one of 16 bolometers) and the multifrequency
Wiener-filtered Maxima-I map, its 124 square degrees drawn to
scale, are shown in the bottom figure. Only the 440 square degrees
within the central rectangle of the entire 1800 square degrees
covered by Boomerang were used in the analysis. \normalsize }
\label{fig:CLdatmap}
\end{figure}

We are only at the beginning of the high precision  CMB era for
primary anisotropies  heralded by the arrival of Boomerang and
Maxima, with interferometers taking data (VSA, CBI, DASI), the
single dish ACBAR about to, and new LDBs to fly in the next few
years (Arkeops, Tophat, Beast/Boost), as well as Boomerang-2001
and the neo-Maxima Maxipol, both concentrating on polarization. In
April 2001, NASA's HEMT-based MAP satellite will launch, with
$12^\prime$ resolution, and in 2007, ESA's bolometer+HEMT-based
Planck satellite is scheduled for launch, with $5^\prime$
resolution.

\vskip 10pt \noindent {\bf The CMB Analysis Pipeline:} Analyzing
Boomerang and other experiments involves a pipeline that takes (1)
the timestream in each of the bolometer channels coming from the
balloon plus information on where it is pointing and turns it into
(2) spatial maps for each frequency characterized by average
temperature fluctuation  values in each pixel (Fig.~1) and a
pixel-pixel correlation matrix characterizing the noise, from
which various statistical quantities are derived, in particular
(3) the temperature power spectrum as a function of multipole
(Fig.~1), grouped into bands, and two band-band error matrices
which together determine the full likelihood distribution of the
bandpowers (Bond, Jaffe \& Knox 1998 [BJK98], BJK00). Fundamental
to the first step is the extraction of the sky signal from the
noise, using the only information we have, the pointing matrix
mapping a bit in time onto a pixel position on the sky. To compare
the data with millions of cosmological models, as we wish to do
here, the radical compression step from 2 to 3 is essential, and
hinges upon an accurate representation of the likelihood surface.

There is generally another step in between (2) and (3), namely
separating the multifrequency spatial maps into the physical
components on the sky: the primary CMB, the thermal and kinematic
Sunyaev-Zeldovich effects, the dust, synchrotron and
bremsstrahlung Galactic signals, the extragalactic radio and
submillimetre sources. The strong agreement among the Boomerang
maps indicates that to first order we can ignore this step, but it
has to be taken into account as the precision increases.  The
Fig.~1 map is consistent with a Gaussian distribution, thus fully
characterized by just the power spectrum. Higher order
(concentration) statistics (3,4-point functions, \etc) tell us of
non-Gaussian aspects, necessarily expected from the Galactic
foreground and extragalactic source signals, but possible even in
the early Universe fluctuations. For example, though
non-Gaussianity occurs only in the more baroque inflation models
of quantum noise, it is a necessary outcome of defect-driven
models of structure formation. (Peaks compatible with Fig.~1 do
not appear in non-baroque defect models, which now appear
unlikely.) Though great strides have been made in the analysis of
Boomerang and Maxima, there is intense needed effort worldwide now
to develop new fast algorithms to deal with the looming megapixel
datasets of LDBs and the satellites (\eg Bond \et 1999, Szapudi
\et 2000).

\section{Cosmic Parameters}

\noindent
 {\bf Parameters of Structure Formation:}
 We usually adopt the restricted set of 7 cosmological parameters used in
Let00 and Jet00,
$\{\Omega_\Lambda,\Omega_{k},\omega_b,\omega_{cdm}, n_s,\tau_C,
\sigma_8\}$. The curvature energy is $\Omega_k \equiv
1-\Omega_{tot}$. The dark energy parameterized here by
$\Omega_\Lambda$ could have complex dynamics associated with it,
\eg if it is the energy density of a scalar field which dominates
at late times (now often termed a quintessence field, $Q$, with
energy $\Omega_Q$, \eg Steinhardt 2000).  One popular
phenomenology is to add one more parameter, $w_Q = p_Q /\rho_Q$,
where $p_Q$ and $\rho_Q$ are the pressure and density of the
$Q$-field. Thus $w_Q=-1$ and $\Omega_Q=\Omega_\Lambda$ for the
cosmological constant. We have also allowed $w_Q$ to float.

We use 2 parameters to characterize the early universe primordial
power spectrum of gravitational potential fluctuations $\Phi$, one
giving the overall power spectrum amplitude ${\cal
P}_{\Phi}(k_n)$,  and one defining the shape, a spectral tilt $n_s
(k_n) \equiv 1+d\ln {\cal P}_{\Phi}/d \ln k$, both at some
(comoving) normalization wavenumber $k_n$. We really need another
2,  ${\cal P}_{GW}(k_n)$ and $n_t(k_n)$, associated with the
gravitational wave component. In inflation, the amplitude ratio is
related to $n_t$ to lowest order, with ${\cal O}(n_s-n_t)$
corrections  at higher order, \eg B96. There are also useful
limiting cases for the $n_s-n_t$  relation. However, as one allows
the baroqueness of the inflation models to increase, one can
entertain a plethora of power spectra (with fully $k$-dependent
$n_s(k)$ and $n_t(k)$) if one is artful enough in designing
inflaton potential surfaces. As well, one can have more types of
modes present, \eg scalar isocurvature modes (${\cal
P}_{is}(k_n),n_{is}(k)$) in addition to, or in place of, the
scalar curvature modes (${\cal P}_{\Phi}(k_n),n_{s}(k)$). However,
our philosophy is consider minimal models first, then see how
progressive relaxation of the constraints on the inflation models,
at the expense of increasing baroqueness,  causes the parameter
errors to open up. For example, with COBE-DMR and Boomerang, we
can probe the GW contribution, but the data are not powerful
enough to determine much. Planck can in principle probe the
gravity wave contribution reasonably well.

We use another 2 parameters to characterize the transport of the
radiation through the era of photon decoupling, which is sensitive
to the physical density of the various species of particles
present then, $\omega_j \equiv \Omega_j {\rm h}^2$. We really need
4: $\omega_b$ for the baryons, $\omega_{cdm}$ for the cold dark
matter, $\omega_{hdm}$ for the hot dark matter (massive but light
neutrinos), and $\omega_{er}$ for the relativistic particles
present at that time (photons, very light neutrinos, and possibly
weakly interacting products of late time particle decays). For
simplicity, though, we restrict ourselves to the conventional 3
species of  relativistic neutrinos plus photons, with
$\omega_{er}$ therefore fixed by the CMB temperature and the
relationship between the neutrino and photon temperatures
determined by the extra photon entropy accompanying $e^+ e^- $
annihilation. Of particular importance for the pattern of the
radiation is the (comoving) distance sound can have influenced by
recombination (at redshift $z_r= a_r^{-1}-1$),
\begin{equation}
r_s = {6000 \over \sqrt{3}} \mpc \int_{0}^{\sqrt{a_r}} {d\sqrt{a}
\over (\omega_m + \omega_{er} a^{-1})^{1/2} (1+ \omega_b
a/(4\omega_\gamma /3))^{1/2}} \ , \label{eq:soundhor}
\end{equation}
where $\omega_\gamma =
2.46 \times 10^{-5}$ is the photon density, $\omega_{er} = 1.68
\omega_\gamma$ for 3 species of massless neutrinos and $\omega_m
\equiv \omega_{hdm}+\omega_{cdm}+\omega_b$.

 The angular diameter distance is
\begin{eqnarray}
&&{\cal R} =\{d_k {\rm sinh} (\chi_r/d_k), \chi_r, d_k {\rm sin}
(\chi_r/d_k)\}, {\rm where} \, d_k ={3000
\over \sqrt{ |\omega_k|}} \mpc \, , \label{eq:angdiamdist}\\
&& \chi_r = 6000 \mpc \int_{\sqrt{a_r}}^{1} { d\sqrt{a}\over
(\omega_m + \omega_Q a^{-6w_Q} +\omega_k a)^{1/2}} \, . \nonumber
\end{eqnarray}
The 3 cases are for negative, zero and positive mean curvature,
$d_k$ is the curvature scale and $\chi_r$  is the comoving
distance to recombination. The location of the first acoustic
peak,  $L_{Pk} \approx 0.746\pi  {\cal R}/r_s$ (\eg Efstathiou and
Bond 1999, hereafter EB99), depends upon $\omega_b$ through the
sound speed as well as on $\omega_{k}$, $\omega_\Lambda$ and
$\omega_m$. Thus $L_{Pk}$ defines a functional relationship among
these parameters, a {\it degeneracy} (EB99) that would be exact
except for the integrated Sachs-Wolfe effect, associated with the
change of $\Phi$ with time if $\Omega_\Lambda$ or $\Omega_k$ is
nonzero.

 Our 7th parameter is an astrophysical one, the Compton "optical depth"
$\tau_C$ from a reionization redshift $z_{reh}$ to the present. It
lowers ${\cal C}_\ell$ by $\exp(-2\tau_C)$ at the high $\ell$'s
probed by Boomerang. For typical models of hierarchical structure
formation, we expect $\tau_C \lta 0.2$. It is partly degenerate
with $\sigma_8$ and cannot be determined at this precision by CMB
data now.

The LSS also depends upon our parameter set. Here we use a set of
(relatively weak) constraints on $\ln \sigma_8^2$ from cluster
abundance data and on $\Gamma + (n_s-1)/2$ from galaxy clustering
data (B96, Let00). $\sigma_8^2$ is a bandpower for density
fluctuations on a scale associated with rare clusters of galaxies,
$8\hmpc$, which we often use in place of ${\cal P}_\Phi (k_n)$ or
${\cal C}_{10}$ for the amplitude parameter. The
mass-density-power-spectrum-shape-parameter $\Gamma$ depends upon
$\{\omega_m, \omega_{er}, \omega_b, {\rm h}\}$ and is related to
the horizon scale when the energy density in relativistic
particles equals that in nonrelativistic ones.

When we allow for freedom in $\omega_{er}$, the abundance of
primordial helium, tilts of tilts ($dn_{\{s,is,t\}}(k_n)/d\ln k,
...$) for 3 types of perturbations, the parameter count would be
17, and many more if we open up full theoretical freedom in
spectral shapes. However, as we shall see, as of now only 3 or 4
combinations can be determined with 10\% accuracy with the CMB.
Thus choosing 7 is adequate for the present, 6 of which are
discretely sampled, with generous boundaries.\footnote{The
specific discrete parameter values used for the ${\cal
C}_\ell$-database in this analysis were: ($\Omega_\Lambda =$
0,.1,.2,.3,.4,.5,.6,.7,.8,.9,1.0,1.1), ($\Omega_k =$
.9,.7,.5,.3,.2,.15,.1,.05,0,-.05,-.1,-.15,-.2,-.3,-.5), ($\tau_c
=$0, .025, .05, .075, .1, .15, .2, .3, .5) when $w_Q=$ --1, with
slightly different ranges (and $\Omega_{tot}$=1) when we allow
$w_Q$ to float; ($\omega_c =$ .03, .06, .12, .17, .22, .27, .33,
.40, .55, .8), ($\omega_b =$ .003125, .00625, .0125, .0175, .020,
.025, .030, .035, .04, .05, .075, .10, .15, .2), ($n_s = $1.5,
1.45, 1.4, 1.35, 1.3, 1.25, 1.2, 1.175, 1.15, 1.125, 1.1, 1.075,
1.05, 1.025, 1.0, .975, .95, .925, .9, .875, .85, .825, .8, .775,
.75, .725, .7, .65, .6, .55, .5), $\sigma_8^2$ was continuous, and
there were 4 experimental parameters for Boomerang and Maxima
(calibration and beam uncertainties), as well as other calibration
parameters for some of the prior-CMB experiments.} For drawing
cosmological conclusions we adopt a weak prior probability on the
Hubble parameter and age: we restrict ${\rm h} $ to lie in the
0.45 to 0.9 range, and the age to be above 10 Gyr.

\vskip 10pt \noindent {\bf The First Peak, $\Omega_{tot}$ and
$\Omega_\Lambda$:}  $L_{Pk}$ and its errors are found from the
average and variance of $\ln L_{Pk}$, taken {\it wrt} the full
probability function over our database described above (restricted
for this exercise to the $\tau_C=0$ part and with the weak prior).
As more CMB data were added, $L_{Pk}$ evolved and the errors
shrunk considerably: $240^{+70}_{-54}$ for April 99 data,
$220^{+30}_{-27}$ for TOCO+4.99 data, $224^{+23}_{-21}$ when
Boomerang-NA was added, and $212^{+7}_{-7}$ when Boomerang-LDB and
Maxima-1 were added to prior-CMB. The latter contrasts with
$202^{+7}_{-7}$ for Boomerang-LDB alone, $226^{+17}_{-16}$ for
Maxima-1 alone, and $208^{+7}_{-7}$ for the combination. The
numbers change a bit depending upon exactly what prior one chooses
or what functional forms one averages over. For the choice here,
although there is a large difference in the mean between the
Maxima and Boomerang numbers, it is not unreasonable within the
errors. Other ways of doing this make the discrepancy seem more
statistically significant (\eg Page 2000, these proceedings).

 In Fig.~2,  we show the lines of constant $L_{Pk}\propto {\cal R}/r_s$ in the
$\Omega_{tot}$--$\Omega_\Lambda$ plane, for given $\omega_m$ and
$\omega_b$, using the formulas given above and discussed in more
detail in  EB99.  The $\pm 10$ band around 210 corresponds to our
best $L_{Pk}$ estimate of $212\pm 7$ using all current CMB data.
Note that the constant $L_{Pk}$ lines look rather similar to the
contours shown in the right panel, showing that the ${\cal R}/r_s$
degeneracy plays a large role in determining the contours. The
contours hug the $\Omega_{tot}=1$ line more closely than the
allowed $L_{Pk}$ band does for the maximum probability values of
$\omega_m$ and $\omega_b$, because of the shift in the allowed
$L_{Pk}$ band as $\omega_m$ and $\omega_b$ vary in this plane. See
also Bond \et (2001b, [capp2K]) for $L_{Pk}$ plots in the
quintessence plane, $w_Q$-$\Omega_Q$, which demonstrate why $w_Q$
is poorly determined by CMB alone.

\begin{table}
\caption{\small Cosmological parameter values and their 1-sigma
errors are shown, determined after marginalizing over the other
$6$ cosmological and $4^{+}$ experimental parameters, for
B98+Maxima-I+prior-CMB and the weak prior, $0.45 \le {\rm h} \le
0.9$, age $> 10$ Gyr. The LSS prior was also designed to be weak.
In the first set $\Omega_{tot}$ varies, in the second set it is
fixed to unity. Similar tables for B98+DMR are given in Let00 and
for B98+MAXIMA-I+DMR in Jet00. We have set the quintessence
$w_Q=p_Q/\rho_Q$ parameter to $-1$, the cosmological constant
case, but the last line shows the limit on $w_Q$ if we allow it to
vary (the other parameters do not move much). SN1 results on $w_Q$
with earlier CMB data were given in Perlmutter \et (1999). The
detections in the table are clearly very stable if extra "prior"
probabilities for LSS and SN1 are included, and are also stable
with much stronger priors on ${\rm h}$, but do move if the
BBN-derived $0.019 \pm 0.002$ prior is imposed. If $\Omega_{tot}$
is varied, parameters derived from our basic 7 come out to be:
age=$13.2\pm 1.3$ Gyr, ${\rm h}=0.70 \pm 0.09$, $\Omega_m=0.35\pm
.06$, $\Omega_b=0.065 \pm .019$. Restriction to $\Omega_{tot}=1$
yields: age=$11.6\pm 0.4$ Gyr, ${\rm h}=0.80\pm .04$,
$\Omega_m=0.31\pm .03$, $\Omega_b=0.046 \pm .005$. \normalsize}
\label{tab:exptparams}
\begin{center}
\hspace{-0.15truecm}
\begin{tabular}{|l|llll|}
\hline
   & cmb & +LSS & +SN1 &  +SN1+LSS \\
\hline
$\Omega_{tot}$           & $1.09^{+.07}_{-.07}$ &
$1.08^{+.06}_{-.06}$ & $1.04^{+.06}_{-.05}$ & $1.04^{+.05}_{-.04}$ \\
$\Omega_b{\rm h}^2$             & $.031^{+.005}_{-.005}$ &
$.031^{+.005}_{-.005}$ & $.031^{+.005}_{-.005}$ &
$.031^{+.005}_{-.005}$  \\
$\Omega_{cdm}{\rm h}^2$ & $.17^{+.06}_{-.05}$ &
$.14^{+.03}_{-.02}$ & $.13^{+.05}_{-.05}$ & $.15^{+.03}_{-.02}$  \\
$n_s$            & $1.05^{+.09}_{-.08}$ & $1.04^{+.09}_{-.08}$ &
$1.05^{+.10}_{-.09}$ & $1.06^{+.08}_{-.08}$ \\
$\Omega_{\Lambda}$ & $0.48^{+.20}_{-.26}$ & $0.63^{+.08}_{-.09}$ &
$0.72^{+.07}_{-.07}$ & $0.70^{+.04}_{-.05}$  \\
\hline
 & $\Omega_{tot}$  & =1 & CASE & ($w_Q$=-1)  \\
\hline $\Omega_b{\rm h}^2$& $.030^{+.004}_{-.004}$ &
$.030^{+.003}_{-.004}$ & $.030^{+.004}_{-.004}$ &
$.030^{+.003}_{-.004}$ \\
$\Omega_{cdm}{\rm h}^2$& $.19^{+.06}_{-.05}$ & $.17^{+.02}_{-.02}$
& $.16^{+.03}_{-.03}$ & $.17^{+.01}_{-.02}$\\
$n_s$ & $1.02^{+.08}_{-.07}$ & $1.03^{+.08}_{-.07}$ &
$1.03^{+.08}_{-.07}$ & $1.04^{+.07}_{-.07}$  \\
$\Omega_{\Lambda}$& $0.58^{+.17}_{-.27}$ & $0.66^{+.04}_{-.06}$ &
$0.71^{+.06}_{-.07}$ & $0.69^{+.03}_{-.05}$ \\
\hline
$w_{Q}$ (95\%)& $< -0.29$ & $<-0.33$ & $< -0.69$ & $<-0.73$  \\
\hline
\end{tabular}
\end{center}
\end{table}

\begin{figure}
\centerline{\hspace{5pt}\epsfxsize=3.5in\epsfbox{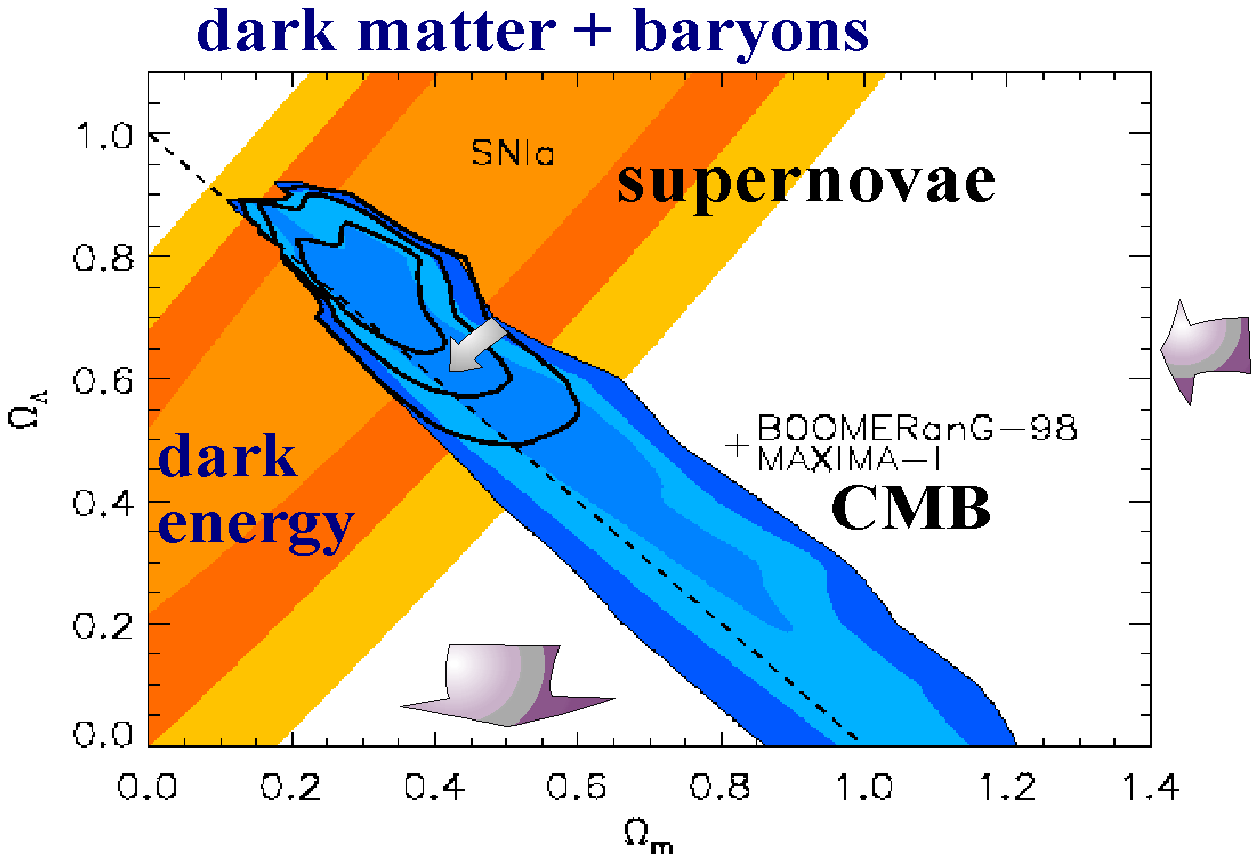}
\hspace{-15pt}\epsfxsize=3.5in\epsfbox{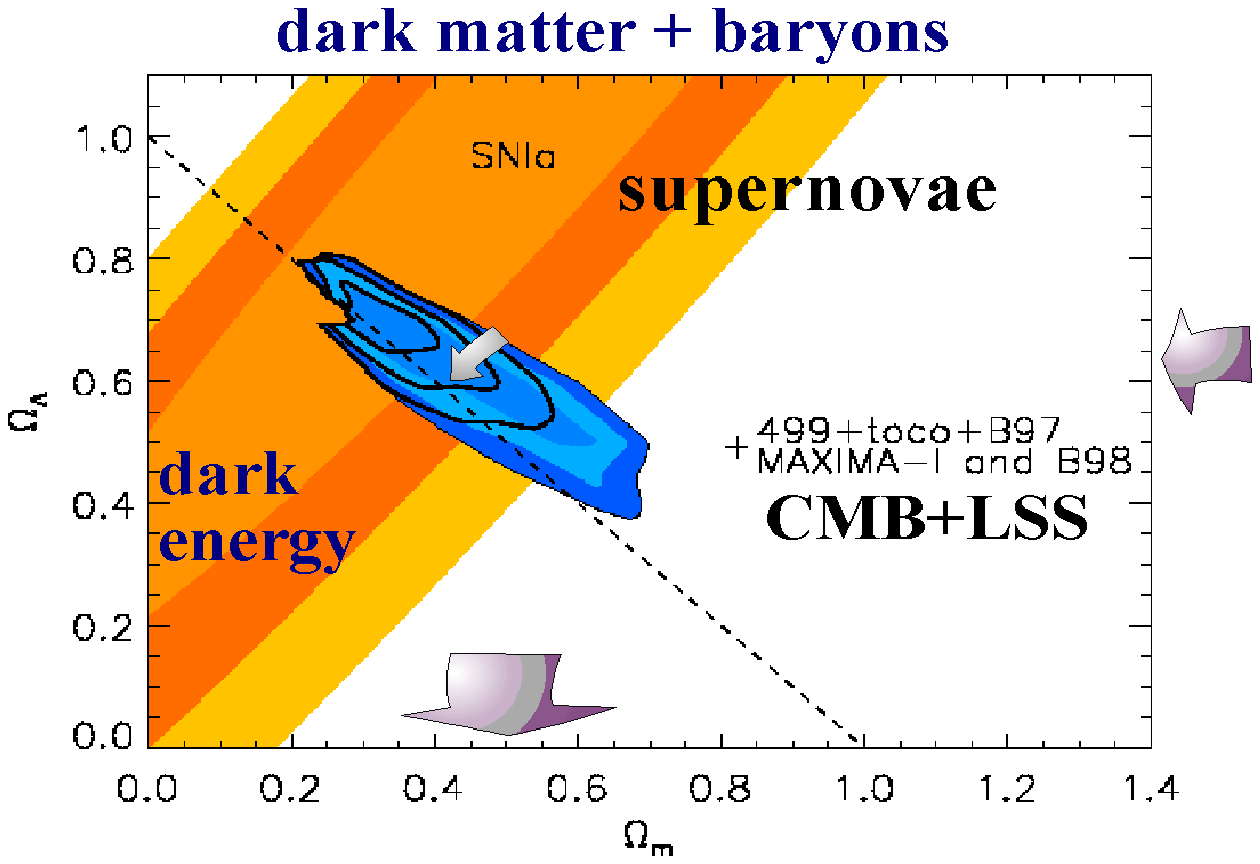}}
\vspace{-20pt}
\centerline{\hspace{5pt}\epsfxsize=3.0in\epsfbox{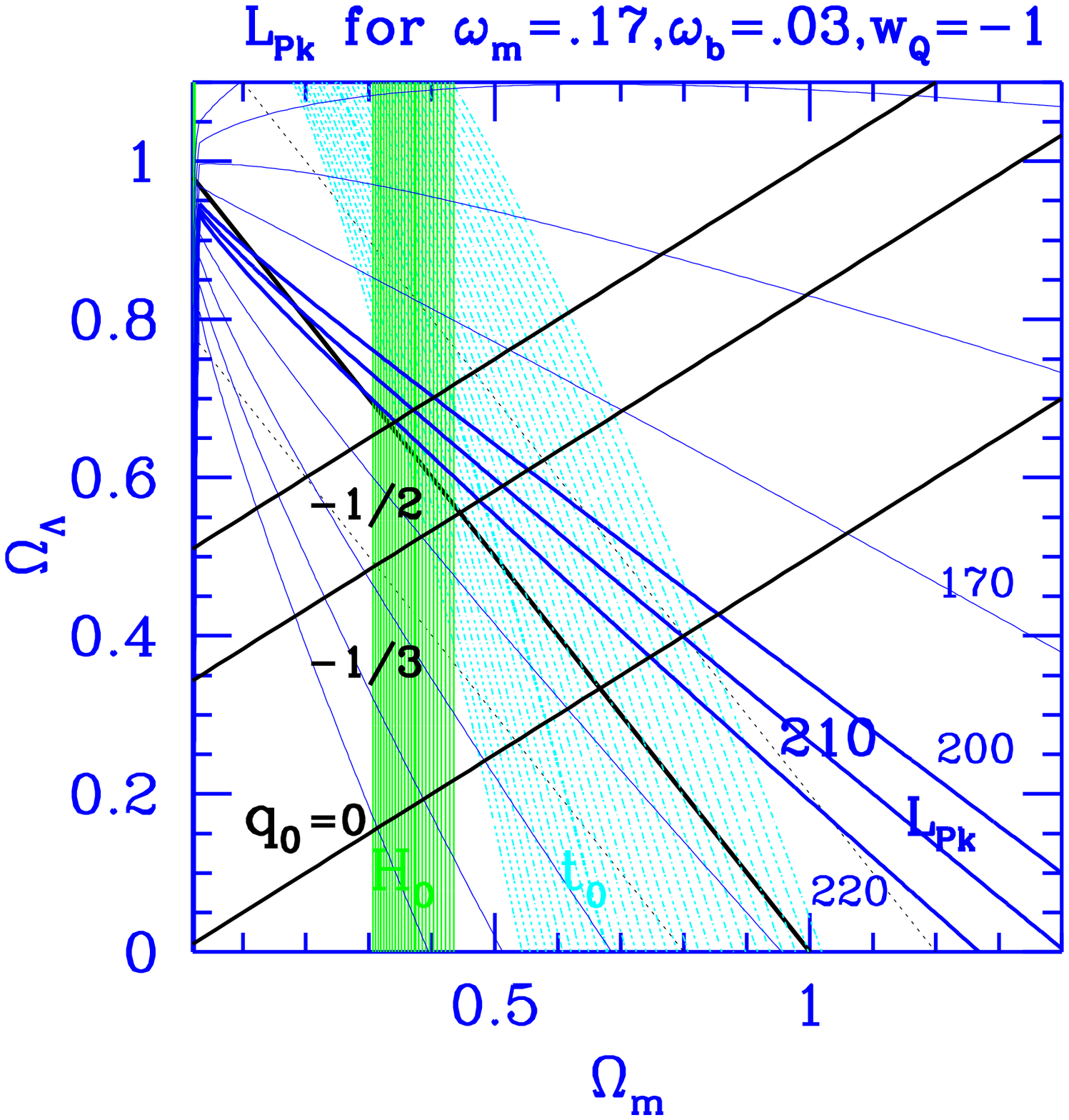}
 \hspace{-25pt}\epsfxsize=3.0in\epsfbox{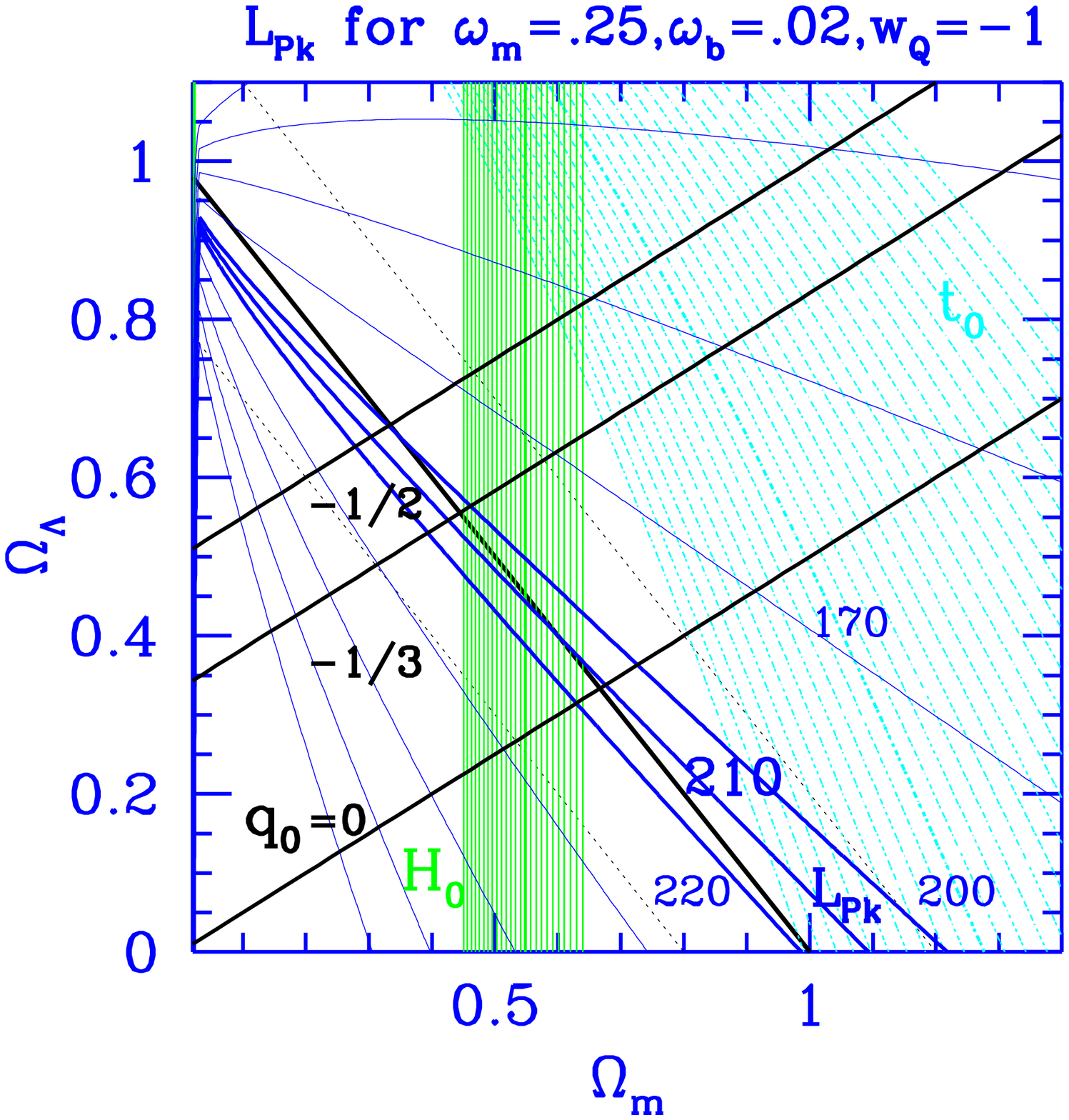}}
 \vspace{-10pt} \caption{\small The top panels show
1,2,3-sigma likelihood contours for the weak-H+age prior
probability (right) and when the LSS "prior" is included (left).
In the right panel, "prior-CMB" experiments
(TOCO+Boomerang-NA+"April 99") were included instead of just DMR,
but the figure is very similar for Boomerang+Maxima+DMR. The
supernova contours are also plotted, and the solid contour lines
are what you get when you combine the two likelihoods.  The bottom
panels show lines of constant $L_{Pk}$ in the
$\Omega_m$--$\Omega_\Lambda$ plane for two choices of $\{ \omega_m
,\omega_b \}$, left the most probable values, right when the
current BBN constraint is imposed (lowering $\omega_b$ increases
the sound speed, decreasing $L_{Pk}$, and varying $\omega_m$ also
shifts it). The $0.65 < {\rm h}< 0.75 $ (heavier shading, $H_0$)
and $11< $ age $< 15$ (lighter shading, $t_0$) ranges and
decelerations $q_0=0,-1/3,-1/2$ are also noted. The sweeping back
of the $L_{Pk}$ curves into the closed models as $\Omega_\Lambda$
is lowered shows that even if $\Omega_{tot}$=1 is the correct
answer, the phase space results in a 1D projection onto the
$\Omega_{tot}$ axis that would be skewed to $\Omega_{tot}>1$, a
situation we see in Table~1. Note that the contours in the top
left panel are near the diagonal $\Omega_{tot}=1$ line, but also
follow a weighted average of $L_{Pk}\sim 210$ lines. This
approximate degeneracy implies $\Omega_\Lambda$ is poorly
constrained for CMB-only, but it is broken when LSS is added,
giving a solid SN1-independent $\Omega_\Lambda$ "detection".
$L_{Pk}$ and contour plots in the $w_Q$--$\Omega_Q$ plane for
$\Omega_{tot}$=1 are given in capp2K. \normalsize }
\label{fig:OmOL}
\end{figure}

\vskip 10pt \noindent {\bf Marginalized Estimates of our Basic 7
Parameters:} Table~1 shows there are strong detections with only
the CMB data for $\Omega_{tot}$, $\omega_b$ and $n_s$ in the
minimal inflation-based 7 parameter set, and a reasonable
detection of $\omega_{cdm}$. The ranges quoted are Bayesian 50\%
values and the errors are 1-sigma, obtained after projecting
(marginalizing) over all other parameters. That $\Omega_\Lambda$
is not well determined is a manifestation of the
$\Omega_{tot}$--$\Omega_\Lambda$ near-degeneracy discussed above,
which is broken when LSS is added because the CMB-normalized
$\sigma_8$ is quite different for open cf. $\Lambda$ models.
Supernova at high redshift give complementary information to the
CMB, but with CMB+LSS (and the inflation-based paradigm) we do not
need it: the CMB+SN1 and CMB+LSS numbers are quite compatible. In
our space, the Hubble parameter, ${\rm h}= (\sum_j (\Omega_j{\rm
h}^2 ))^{1/2}$, and the age of the Universe, $t_0$, are derived
functions of the $\Omega_j{\rm h}^2$: representative values are
given in the Table caption.

Fig.~3 shows  how the parameter estimations evolved as more CMB
data were added (for the weak+LSS prior). With just the
COBE-DMR+LSS data, the 2-sigma contours were already localized in
$\omega_{cdm}$. Without LSS, it took the addition of Maxima-1
before it began to localize. $\Omega_k$ localized near zero when
TOCO was added to the April 99 data, more so when Boomerang-NA was
added, and much more so when Boomerang-LDB and Maxima-1 were
added. Some $n_s$ localization occurred with just "prior-CMB"
data. $\omega_b$ really focussed in with Boomerang-LDB and
Maxima-1, as did $\Omega_\Lambda$.

\begin{figure}
\centerline{\epsfig{file=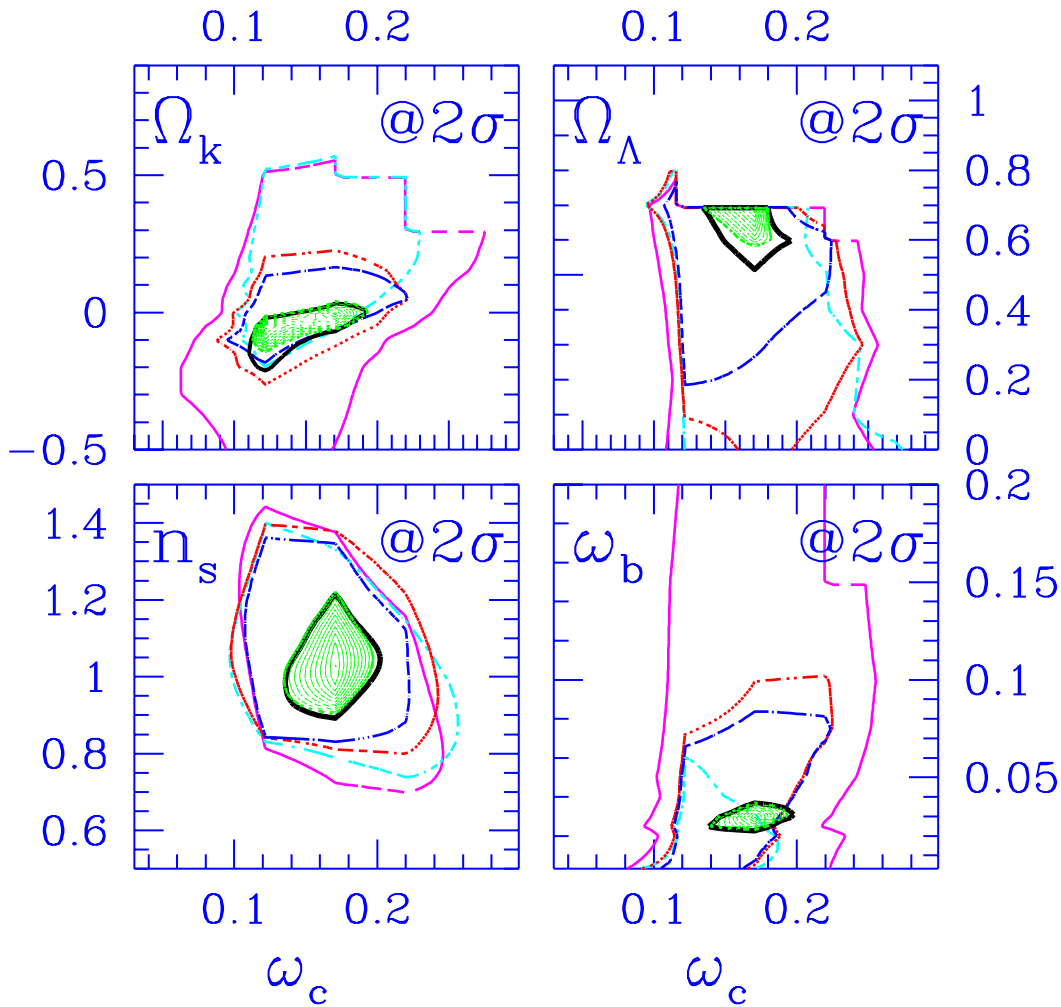,height=6.0in}}
\vspace{-10pt} \caption{\small 2-sigma likelihood contours for the
dark matter density $\omega_c =\Omega_{cdm}{\rm h}^2$ and
$\{\Omega_k,\Omega_\Lambda,n_s,\omega_b\}$ are plotted for LSS,
the weak-H+age cosmological prior,   and the following CMB
experimental combinations: DMR (short-dash); the "April 99" data
(short-dash long-dash); TOCO+4.99 data (dot short-dash);
Boomerang-NA+TOCO+4.99 data (dot long-dash, termed "prior-CMB");
Boomerang-LDB+Maxima-1+Boomerang-NA+TOCO+4.99 data (heavy solid,
all-CMB). These $2\sigma$ lines tend to go from outside to inside
as more CMB experiments are added. The smallest
 2-sigma region (dotted and interior) shows SN1+LSS+all-CMB, when
SNI data is added. For the $\Omega_\Lambda$, $n_s$ and $\omega_b$
plots , $\Omega_{tot}$=1 has been assumed, but the values do not
change that much if $\Omega_{tot}$ floats, as Table~1 shows for
all-CMB. (The discreteness of the ${\cal C}_\ell$--database is
responsible for the sharpness of the contour edges, especially
evident in the $\omega_c$--$\Omega_\Lambda$ CMB+LSS case. Though
different interpolation schemes can round these off somewhat, it
is the price of our finite grid.) \normalsize }
\label{fig:cmbLSS2sig}
\end{figure}

 We have also considered what happens as we let
$\Omega_{m\nu}/\Omega_m$, the fraction of the matter in massive
neutrinos,  vary from 0 to 0.3, for LSS + all of the CMB data and
$\Omega_{tot}$=1  (Bond \et 2001a, [$\nu$2K]). Until Planck
precision, the CMB data by itself will not be able to strongly
discriminate this ratio. Adding HDM does have a strong impact on
the CMB-normalized $\sigma_8$ and the shape of the density power
spectrum (effective $\Gamma$ parameter), both of which mean that
when LSS is included, adding some HDM to CDM is strongly preferred
in the absence of $\Omega_\Lambda$. However, though higher
$\Omega_m$ is preferred at the expense of less dark energy,
significant $\Omega_\Lambda$ is still required (see $\nu$2K for
the evolution of the CMB+LSS 2-sigma contours in the
$(\omega_{hdm}+\omega_{cdm})$--$\Omega_\Lambda$ plane as
$\Omega_{m\nu}/\Omega_m$ is varied). The $\omega_b$ and $n_s$
likelihood curves are essentially independent of
$\Omega_{m\nu}/\Omega_m$.

\vskip 10pt \noindent {\bf The Future, Forecasts for Parameter
Eigenmodes:} We can also forecast dramatically improved precision
with further analysis of Boomerang and Maxima, future LDBs, MAP
and Planck. Because there are correlations among the physical
variables we wish to determine, including a number of
near-degeneracies beyond that for $\Omega_{tot}$--$\Omega_\Lambda$
(EB99), it is useful to disentangle them, by making combinations
which diagonalize the error correlation matrix, "parameter
eigenmodes" (\eg B96, EB99). For this exercise, we will add
$\omega_{hdm}$ and $n_t$ to our parameter mix, but set $w_Q$=$-1$,
making 9. (The ratio ${\cal P}_{GW}(k_n)/{\cal P}_\Phi (k_n)$ is
treated as fixed by $n_t$, a reasonably accurate inflation theory
result.) The forecast for Boomerang+DMR based on the 440 square
degree patch with a single 150 GHz bolometer used in the published
data is 3 out of 9 linear combinations should be determined to
$\pm 0.1$ accuracy. This is indeed what we get in the full
analysis of Let00 for CMB only. If 4 of the 6 150 GHz channels are
used and the region is doubled in size, we predict 4/9 could be
determined to $\pm 0.1$ accuracy. The Boomerang team is still
working on the data to realize this promise. And if the optimistic
case for all the proposed LDBs is assumed, 6/9 parameter
combinations could be determined to $\pm 0.1$ accuracy, 2/9 to
$\pm 0.01$ accuracy. The situation improves for the satellite
experiments: for MAP, we forecast  6/9 combos to $\pm 0.1$
accuracy, 3/9 to $\pm 0.01$ accuracy; for Planck,  7/9 to $\pm
0.1$ accuracy, 5/9 to $\pm 0.01$ accuracy. While we can expect
systematic errors to loom as the real arbiter of accuracy, the
clear forecast is for a very rosy decade of high precision CMB
cosmology that we are now fully into.

\end{document}